\documentclass[journal]{IEEEtran}
\usepackage{latexsym}
\usepackage{graphics}
\usepackage{epstopdf}
\usepackage{epsfig}
\usepackage{amsfonts}
\usepackage{float}
\usepackage{amssymb}
\usepackage{stfloats}
\usepackage{multirow}
\usepackage{multicol}
\usepackage{amsmath}
\usepackage{bm}
\usepackage{array}
\usepackage{cite}
\usepackage{tabularx}
\usepackage{enumerate}
\usepackage{times}
\usepackage[noend]{algpseudocode}
\usepackage{threeparttable}
\usepackage{amsfonts, amssymb, amsthm}
\usepackage{latexsym, graphicx}
\usepackage{epsfig}
\usepackage{float}
\usepackage{fancyhdr}
\usepackage{subfigure}
\usepackage{graphicx}
\usepackage{color}
\usepackage{amssymb}
\usepackage{amsfonts}
\usepackage{subfigure}
\usepackage{balance}
\usepackage{tabularx}
\usepackage{bm}
\usepackage{multirow}
\usepackage{graphicx,times,amsmath}
\usepackage{amsfonts}
\usepackage{mathrsfs}
\usepackage[table]{xcolor}

\usepackage{amsfonts}
\usepackage{booktabs}
\usepackage{siunitx}

\usepackage{algorithm2e}
\RestyleAlgo{ruled}

\ifCLASSINFOpdf

\else

\fi

\begin{document}
\title{Physics-Informed Kolmogorov-Arnold Networks for Power System Dynamics}

\author{Hang~Shuai,~\IEEEmembership{Member,~IEEE,}
        and~Fangxing~Li,~\IEEEmembership{Fellow,~IEEE}
\thanks{H. Shuai and F. Li are with the Department of Electrical Engineering and Computer Science, University of Tennessee, Knoxville, TN, 37996, USA. (e-mail: hshuai1@utk.edu; fli6@utk.edu)}
\thanks{This work was supported in part by the CURENT research center.}
\vspace{-2em}
}

\markboth{Shuai H., Li F., Physics-Informed Kolmogorov-Arnold Networks...}%
{Shell \MakeLowercase{\textit{et al.}}: Bare Demo of IEEEtran.cls for IEEE Journals}

\maketitle

\begin{abstract}
This paper presents, for the first time, a framework for Kolmogorov-Arnold Networks (KANs) in power system applications. 
Inspired by the recently proposed KAN architecture, this paper proposes physics-informed Kolmogorov-Arnold Networks (PIKANs), a novel KAN-based physics-informed neural network (PINN) tailored to efficiently and accurately learn dynamics within power systems. 
The PIKANs present a promising alternative to conventional Multi-Layer Perceptrons (MLPs) based PINNs, achieving superior accuracy in predicting power system dynamics while employing a smaller network size.
Simulation results on a single-machine infinite bus system and a 4-bus 2-generator system underscore the accuracy of the PIKANs in predicting rotor angle and frequency with fewer learnable parameters than conventional PINNs. 
Furthermore, the simulation results demonstrate PIKANs' capability to accurately identify uncertain inertia and damping coefficients.
This work opens up a range of opportunities for the application of KANs in power systems, enabling efficient determination of grid dynamics and precise parameter identification.

\end{abstract}


\IEEEpeerreviewmaketitle

\section{Introduction}

\IEEEPARstart{D}{eep} learning (DL) has demonstrated remarkable success in addressing complex tasks, particularly in fields where precise mathematical models are difficult to establish, such as computer vision, natural language processing 
, protein structure prediction 
, and medical image analysis \cite{razzak2018deep}. In the power sector, DL has also increasingly been investigated for applications such as renewable energy forecasting \cite{Hong_2020}, fault detection \cite{Jiang2014}, power system stability assessment \cite{Ren2022}, reflecting its growing influence and great application potential in future power grids.

Regarding power system dynamics, significant efforts have been made to develop various data-driven algorithms for the online identification of power system dynamics \cite{Sinha2020, Satheesh2022, bhusal2021deep}.
Among these, DL techniques have been increasingly utilized \cite{zhang2022review}.
However, these DL-based approaches often lacked integration with the underlying power system model. 
As a result, they relied heavily on the quality and quantity of training data, necessitating large datasets and complex neural network architectures. 
Considering this, researchers futher proposed physics-informed neural network (PINN) based algorithms for power system dynamic identification.
For example, in \cite{misyris2020physics}, a PINN approach was developed to learn the rotor angle and frequency dynamics of a single-machine infinite bus (SMIB) power system. The PINN based method leverages the underlying physical model, resulting in significantly reduced computation times and a lesser need for training data.
Researchers further proposed a practical framework for identifying essential features of nonlinear voltage dynamics. This approach converts PINNs into a mixed-integer linear program \cite{Misyris2021capturing}. It enables adjustment of the neural network's output conservativeness concerning stability boundaries, eliminating the necessity for exhaustive time-domain simulations.

Despite promising results in designing PINNs for power system dynamics, there remains significant room for improvement in the accuracy of the learned dynamic models.
For instance, the PINNs agent developed in \cite{misyris2020physics}, exhibits relative $L_2$ errors of 2.37\% between the exact and predicted solutions for the rotor angle of a SMIB power system.
When used to identify the generator inertia constant and the damping coefficient of a power system, the mean error of parameter identification of PINNs would reach around 50\% when only limited measurements (such as rotor angle) are available \cite{stiasny2021physics}.
Furthermore, the aforementioned PINNs agent struggles to effectively learn both the stable and unstable dynamics of the same power system. This limitation necessitates the use of distinct, trained PINNs to achieve high accuracy in stable and unstable regimes \cite{misyris2020physics}.
In this work, inspired by recently proposed Kolmogorov-Arnold Networks (KANs) in \cite{liu2024kan}, we propose physics-informed KANs (PIKANs) algorithm for accurately predicting power system angular and frequency dynamics that reduces dependency on training data and enables a more smaller number of
learnable parameter in neural networks.

Current DL architectures such as deep neural networks (DNNs), recurrent neural networks (RNNs), convolutional neural networks (CNNs) and PINNs, largely rely on Multi-Layer Perceptrons (MLPs) \cite{lecun2015deep}. MLPs are fully connected neural networks featuring fixed activation functions in neurons, while weights associated with network connections are adjusted using backpropagation techniques during training.
Further, universal approximation theorems \cite{chen1995universal, lu2020universal} imply that MLPs are universal approximators, which can represent a wide variety of interesting functions with appropriate weights and activation functions.
However, MLPs based DL techniques face challenges such as the requirement for large training datasets, catastrophic forgetting \cite{kemker2018measuring}, and a lack of interpretability \cite{Fan2021On} (often referred to as the "black box" problem).
KANs \cite{liu2024kan}, promising alternatives to MLPs, also feature fully-connected network structures. 
Unlike MLPs, KANs place learnable activation functions on the edges, which usually allow much smaller computation graphs than MLPs and could reach more accurate learning results at the same time.
While MLPs have the potential to learn generalized additive structures, they often struggle with efficiently approximating exponential and sine functions using traditional activation functions, such as ReLU. 
In contrast, KANs excel in learning both compositional structures and univariate functions effectively, thereby significantly outperforming MLPs \cite{liu2024kan}.
Considering sine and cosine functions are fundametal funcations in power system dynamics models, so KANs would have great potential to be more effective at representing dynamics in power systems than MLPs. 
In summary, KANs are mathematically sound, accurate and interpretable, which offer a range of opportunities in power systems by precisely and adaptively determining grid dynamics as described by differential-algebraic equations (DAEs). 

This is the first work to propose the use of KANs for power system applications. Specifically, we utilize the swing equation in power systems as an example to demonstrate their potential. 
We also demonstrate the proposed method can be used to estimate uncertain inertia and damping coefficients. The main contributions of this work can be summarized as follows:
\begin{itemize}
    \item For the first time, we present a framework that integrates KANs with the PINNs architecture for power system applications, and PIKAN algorithms for power system dynamics are developed. We propose a PIKAN training procedure that leverages the power system swing equation model to reduce data dependency and achieve high accuracy.
\item The performance of the proposed method is demonstrated on a SMIB system and a 4-bus 2-generator system. The simulation results show that PIKANs achieve higher accuracy in solving the DAEs of power systems with smaller neural network size compared to traditional MLP-based PINNs.
\end{itemize}

This paper is organized as follows: Section II introduces the power system dynamic model investigated in this work. Section III presents KANs and the framework designation that integrates KANs with the PINNs architecture for the power system dynamic application.
Section IV shows numerical simulation results on the two testing systems demonstrating the performance of the proposed algorithm.
Section V discusses the findings and limitations of this work.
Section VI concludes the paper.
\vspace{-1em}

\section{Power System Dynamic Model}
Power system dynamics are described by swing equations.
By assuming the bus voltage magnitudes to be 1 per unit (p.u.), and neglecting the reactive power flows, the frequency dynamics of each generator $i$ can be
described by the following equation \cite{Hang2024safe}:

\begin{equation} \label{eq1}
\begin{aligned}
\frac{d{\theta_i}}{dt} &= \omega_i \\
M_i \cdot \frac{{d\omega_i}}{dt} &= P_{m_i} - P_{e_i} - D_i \cdot \omega_i
\end{aligned}
\end{equation}
where $\theta_i$ and $\omega_i$ are the voltage angle and angular frequency of the generator $i$ (also connected to bus $i$), respectively.
$t$ is the time index.
$M_i$ and $D_i$ are the inertia and damping constant of the generator $i$, respectively.
$P_{m_i}$ is the net power injection.
$P_{e_i}$ is the electrical power output (p.u.) of the $i$th generator, which can be calculated by the following equation \cite{Vijay2019}:
\begin{equation} \label{eq2}
P_{e_i} = \sum_{j=1}^n V_i V_j[B_{ij} \cdot sin(\theta_i - \theta_j) + G_{ij} \cdot cos(\theta_i - \theta_j)]
\end{equation}
where $B_{ij}$ and $G_{ij}$ are the susceptance and conductance of the transmission line between bus $i$ and $j$, respectively.
$V_i$ and $V_j$ represent the voltage magnitudes at bus $i$
and bus $j$, respectively.

For transmission systems, when the line reactance $X$ greatly exceeds the resistance $R$, and assuming the bus voltage is 1 p.u., equation (2) can be simplified to:
\begin{equation} \label{eq3}
P_{e_i} = \sum_{j=1}^n B_{ij} \cdot sin(\theta_i - \theta_j)
\end{equation}

For the frequency dependent load $i$, the frequency dynamics in equation (\ref{eq1}) can be simplified to:
\begin{equation} \label{eq4}
\begin{aligned}
P_{m_i} - P_{e_i} - D_i \cdot \omega_i &= 0
\end{aligned}
\end{equation}
where $\omega_i = \frac{d{\theta_i}}{dt}$.

Therefore, the system dynamics can be described by equations (\ref{eq1}) and (\ref{eq4}), which can
be expressed in the form of a
DAE system:
\begin{equation} \label{eq5}
\begin{aligned}
\dot{\textbf{x}}_{sys} &= \textbf{h}(\textbf{x}_{sys}, \textbf{y}, \textbf{p}; \boldsymbol{\lambda}) \\
\textbf{0} &= \textbf{g}(\textbf{x}_{sys}, \textbf{y}, \textbf{p}; \boldsymbol{\lambda}) \\
\boldsymbol{P_m} &\in [\boldsymbol{P_m^{min}}, \boldsymbol{P_m^{max}}], t \in [0, T]
\end{aligned}
\end{equation}
where $\textbf{x}_{sys} = [\boldsymbol{\theta}; \boldsymbol{\omega}]$ is the power system state variables vector. $\textbf{y} = [ \boldsymbol{P_e}]$ is the algebraic variables vector.
$\textbf{p} = [\boldsymbol{P_m}]$ represents the power system input variables.
\boldsymbol{$\lambda = [\textbf{M}; \textbf{D}; \textbf{B}]$} is the parameter of the power system.

In this work, we focus on using PIKANs to learn dynamics described by equation (\ref{eq5}), and identify uncertain inertia and damping parameters in \boldsymbol{$\lambda$}.

\section{Physics-Informed Kolmogorov-Arnold Networks for Power System Dynamics}
\subsection{Kolmogorov-Arnold Networks}
Based on Kolmogorov-Arnold Representation theorem \cite{kolmogorov1961representation},  for a multivariate continuous function $f$ on a bounded domain, it can be represented by a finite composition of continuous functions of a single variable and the binary operation of addition, as given by the following equation:
\begin{equation} \label{eq6}
f(\textbf{x}) = f(x_1, x_2, \cdots, x_k) = \sum_{q=1}^{2k+1} \Phi_q (\sum_{p=1}^k \phi_{q,p} (x_p))
\end{equation}
where $\phi_{q,p}$: $[0, 1] \rightarrow \mathbb{R}$ and $\Phi_q$: $\mathbb{R} \rightarrow \mathbb{R}$.
$\textbf{x}$ is the input vector.
The theorem shows that learning a high-dimensional function $f$ can be boiled down to learning a polynomial number of 1D univariate functions $\phi_{q,p}$ and $\Phi_q$.
Actually, equation (\ref{eq6}) can be treated as a two layer network having a shape of $[n, 2n+1, 1]$, which has two-layer nonlinearities.
However, these 1D functions
can be non-smooth and even fractal, so they may not be learnable in practice \cite{poggio2020theoretical}.
Thus, the theorem was thought to be practically useless in machine learning.

\begin{figure}
    \centering
    \includegraphics[width=81mm]{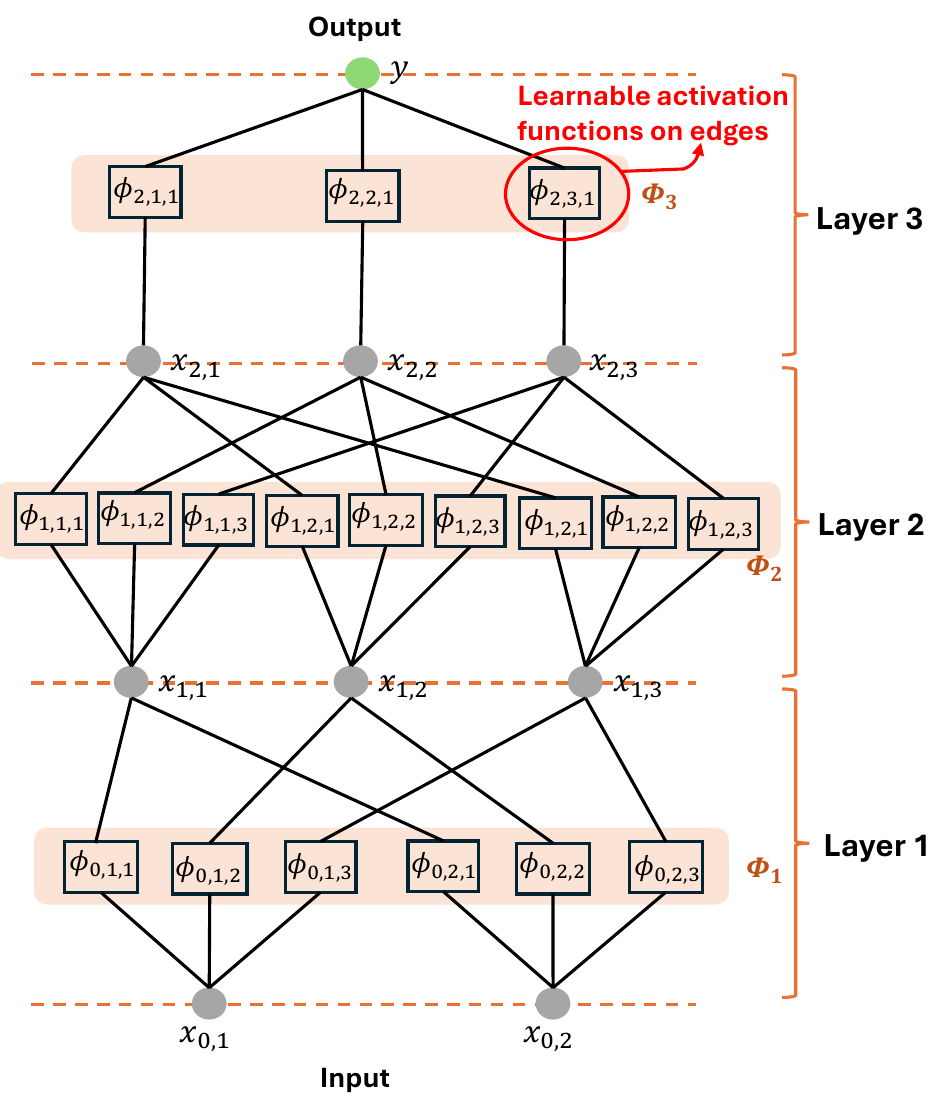}
    \caption{Illustration of a 3-layer KAN having a shape of $[2, 3, 3, 1]$.}
    \label{fig:KAN}
\vspace{-1em}
\end{figure}

To enable the Kolmogorov-Arnold theorem for machine learning, \cite{liu2024kan} innovativaly proposed the KANs architecture, as illustrated in Fig. \ref{fig:KAN}.
In KANs, each 1D function of equation (\ref{eq6}) are parametrized as a B-spline curve. 
Each B-spline curve is with learnable coefficients of local B-spline basis functions. It is worth noting that the activation functions are placed on edges instead of nodes in Fig. \ref{fig:KAN}.
To generalize the network described by equation (\ref{eq6}) to arbitrary widths and depths, \cite{liu2024kan} further defined a KAN layer and stacking more KAN layers as needed. A KAN layer with $n_{l}$-dimensional inputs and $n_{l+1}$-dimensional outputs is defined as a matrix of 1D functions:
\begin{equation} \label{eq7}
\boldsymbol{\Phi}_l = \{\phi_{l, j,i}\}, \ i = 1, 2, \cdots, n_{l}, j = 1, 2, \cdots, n_{l+1}
\end{equation}
where function $\phi_{l,j,i}$ has trainable parameters, which is the activation function that connects the $i^{th}$ neuron in the $l^{th}$ layer and the $j^{th}$ neuron in the $l+1^{th}$ layer.
$l$ is the index of the layer.
Therefore, the output of the $l^{th}$ layer of the KAN is
\begin{align} 
\textbf{x}_{l+1} &= \boldsymbol{\Phi}_l \textbf{x}_l \notag \\
&= \begin{pmatrix}
        \phi_{l,1,1}(\cdot) & \phi_{l,1,2}(\cdot) & \cdots & \phi_{l,1,n_{l}}(\cdot) \\
        \phi_{l,2,1}(\cdot) & \phi_{l,2,2}(\cdot) & \cdots & \phi_{l,2,n_{l}}(\cdot) \\
        \vdots & \vdots & & \vdots \\
        \phi_{l,n_{l+1},1}(\cdot) & \phi_{l,n_{l+1},2}(\cdot) & \cdots & \phi_{l,n_{l+1},n_{l}}(\cdot) \\
\end{pmatrix}
\textbf{x}_{l},
\label{eq8}
\end{align}
In this way, the output of a KAN network composed of $L$ layers can be written as
\begin{equation}\label{eq:9}
    {\rm KAN}(\textbf{x}) = (\boldsymbol{\Phi}_{L-1}\circ \boldsymbol{\Phi}_{L-2}\circ\cdots\circ\boldsymbol{\Phi}_{1}\circ\boldsymbol{\Phi}_{0})\textbf{x}
\end{equation}
where $\textbf{x}\in\mathbb{R}^{n_0}$ is the input vector of the network.
Considering all the above operations are differentiable, KANs can be trained with back propagation techniques.

To make the KAN easy to train, we can design activation functions as given below:
\begin{equation}\label{eq:10}
    \phi(x_{l, i}) = w \cdot (b(x_{l, i}) + spline(x_{l, i}))
\end{equation}
where $w$ is a factor to control the overall magnitude of the activation function. $b(x)$ is a basis function which can be set to
\begin{equation}\label{eq:11}
    b(x_{l, i}) = silu(x_{l, i}) = \frac{x_{l, i}}{1 + e^{-x_{l, i}}}
\end{equation}
$spline(x_{l, i})$ is the spline
function which can be parametrized as a linear combination of B-splines:
\begin{equation}\label{eq:12}
    spline(x_{l, i}) = \sum_s c_s \cdot B_s(x_{l, i})
\end{equation}
where $B_s(x_{l, i})$ is the B-spline function.
During the training process, $spline(\cdot)$ and $w$ are trainable, and we can initialize $spline(\cdot)$ by drawing B-spline coefficients $c_s \sim \mathcal{N}(0, 0.1^2)$ and $w$ initialized according to the Xavier initialization.
It worth noting that other activation functions other than B-spline can be also utilized.
For instance, to address computational cost problem caused by training learnable B-Splines, \cite{bozorgasl2024wav} developed a wavelet KAN architecture based on the work in \cite{liu2024kan}.

For a L-layer KAN with layers of equal width $N$ (which means each layer has $N$ neurons), there are in total $O(N^2L(G+k_b)) \sim O(N^2LG)$ parameters, where $k_b$ and $G$ are the order and intervals of the spline. 
Contrarily, an MLP with depth
$L$ and width $N$ typically requires $O(N^2L)$ parameters, suggesting it might be more parameter-efficient than a KAN. 
However, KANs often operate effectively with much smaller $N$ than MLPs. 
This not only reduces parameter count but also enhances generalization and facilitates interpretability.

\subsection{Physics-informed KANs for Power System Dynamics}
PINNs are universal function approximators that incorporate the knowledge of physical laws governing a given dataset into the neural network training process \cite{raissi2017physics}.
This approach mitigates the need for large amounts of training data and the large network sizes typically required by traditional DNNs.
In PINNs, the architecture consists of a MLP with an input layer, several fully connected hidden layers featuring fixed nonlinear activation functions at each neuron, and an output layer. Each layer transition involves the application of a weight matrix $\boldsymbol{W}_l$ and an activation function $\boldsymbol{\sigma}_l$:
\begin{equation}\label{eq:13}
    {\rm MLP}(\textbf{x}) = (\boldsymbol{W}_{L-1}\circ \boldsymbol{\sigma}_{L-1} \circ\boldsymbol{W}_{L-2}\circ\boldsymbol{\sigma}_{L-2} \circ\cdots\circ\boldsymbol{W}_{1}\circ\boldsymbol{\sigma}_{1} \circ\boldsymbol{W}_{0})\textbf{x}
\end{equation}
During the training process, these weights are adjusted to minimize an objective function, which typically penalizes the difference between the neural network's predictions and the actual labels of the training data.

Based on \cite{raissi2017physics}, the dynamics of a physical system governed by parametrized and nonlinear partial differential equations (PDEs), as shown in equation (\ref{eq:14}), can be effectively learned using PINNs.
\begin{equation}\label{eq:14}
\frac{\partial \textbf{u}}{\partial t} + \mathcal{N} [\textbf{u}; \boldsymbol{\lambda}] = \textbf{0}, \textbf{x} \in \Omega, t \in [0, T]
\end{equation}
where $\textbf{u}(t, \textbf{x})$ is the solution of the PDE, depending on time $t$ and system input $\textbf{x}$.
$\mathcal{N} [\cdot; \boldsymbol{\lambda}]$ is a nonlinear
operator parametrized by $\boldsymbol{\lambda}$.
$\Omega$ is a subset of $\mathbb{R}^D$.
$[0, T]$ is the time interval within which the system evolves.

\begin{figure}
    \centering
    \includegraphics[width=70mm]{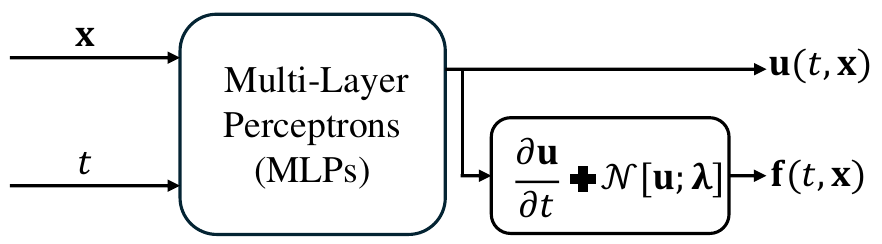}
    \caption{General structure of a PINN \cite{misyris2020physics,raissi2017physics}: it predicts the output $\textbf{u}(t, \textbf{x})$ given inputs $\textbf{x}$ and $t$.}
    \label{fig:PINN}
    \vspace{-1em}
\end{figure}

For the traditional PINNs, we can define a  physics informed neural network $\textbf{f}(t, \textbf{x})$ as equation (\ref{eq:15}) and proceed by approximating $\textbf{u}(t, \textbf{x})$ by a MLP, as illustrated in Fig. \ref{fig:PINN}.
\begin{equation}\label{eq:15}
\textbf{f}(t, \textbf{x}) = \frac{\partial \textbf{u}}{\partial t} + \mathcal{N} [\textbf{u}; \boldsymbol{\lambda}]
\end{equation}
As shown in Fig. \ref{fig:PINN}, the MLPs used for predicting $\textbf{f}(t, \textbf{x})$ shares the same parameters as the MLPs used for predicting $\textbf{u}(t, \textbf{x})$, with the distinction lying in their activation functions. 
The parameters common to both neural networks are optimized by minimizing the following loss function:
\begin{equation}\label{eq:16a}
\begin{aligned}
loss_I &= MSE_u + MSE_f \\
&= \frac{1}{N_u} \sum_{n=1}^{N_u} |\textbf{u}(t_u^n, \textbf{x}_u^n) - \textbf{u}^n|^2 + \frac{1}{N_f} \sum_{n=1}^{N_f} |\textbf{f}(t_f^n, \textbf{x}_f^n)|^2
\end{aligned}
\end{equation}
where loss $MSE_u$ corresponds to the initial and boundary data, while $MSE_f$ enforces the structure imposed by equation (\ref{eq:14}) at a finite set of collocation data points.
The loss $MSE_u$ is calculated over $N_u$ initial and boundary training data points, and $MSE_f$ is calculated over $N_f$ collocation points.
$\textbf{u}(t_u^n, \textbf{x}_u^n)$ and $\textbf{f}(t_f^n, \textbf{x}_f^n)$ are outputs of the PINN, while $\textbf{u}^n$ is the lable value of the $n$th data point.

Considering we can usually obtain the measurement of derivatives of $\textbf{u}(t, \textbf{x})$ with respect to the input $t$, we can also use the following loss function to train the PINN network \cite{stiasny2021physics}:
\begin{equation}\label{eq:16b}
\begin{aligned}
loss_{II} &= MSE_u + MSE_f \\
&= \frac{1}{N_u} \sum_{n=1}^{N_u} |\textbf{u}(t_u^n, \textbf{x}_u^n) - \textbf{u}^n|^2 + |\dot{\textbf{u}}(t_u^n, \textbf{x}_u^n) - \dot{\textbf{u}}^n|^2 \\
&+ \frac{1}{N_f} \sum_{n=1}^{N_f} |\textbf{f}(t_f^n, \textbf{x}_f^n)|^2
\end{aligned}
\end{equation}
By using automatic differentiation in PyTorch, we can easily obtain the derivatives of $\textbf{u}(t_u^n, \textbf{x}_u^n)$ with respect to the input $t$.

\begin{figure*}
    \centering
    \includegraphics[width=120mm]{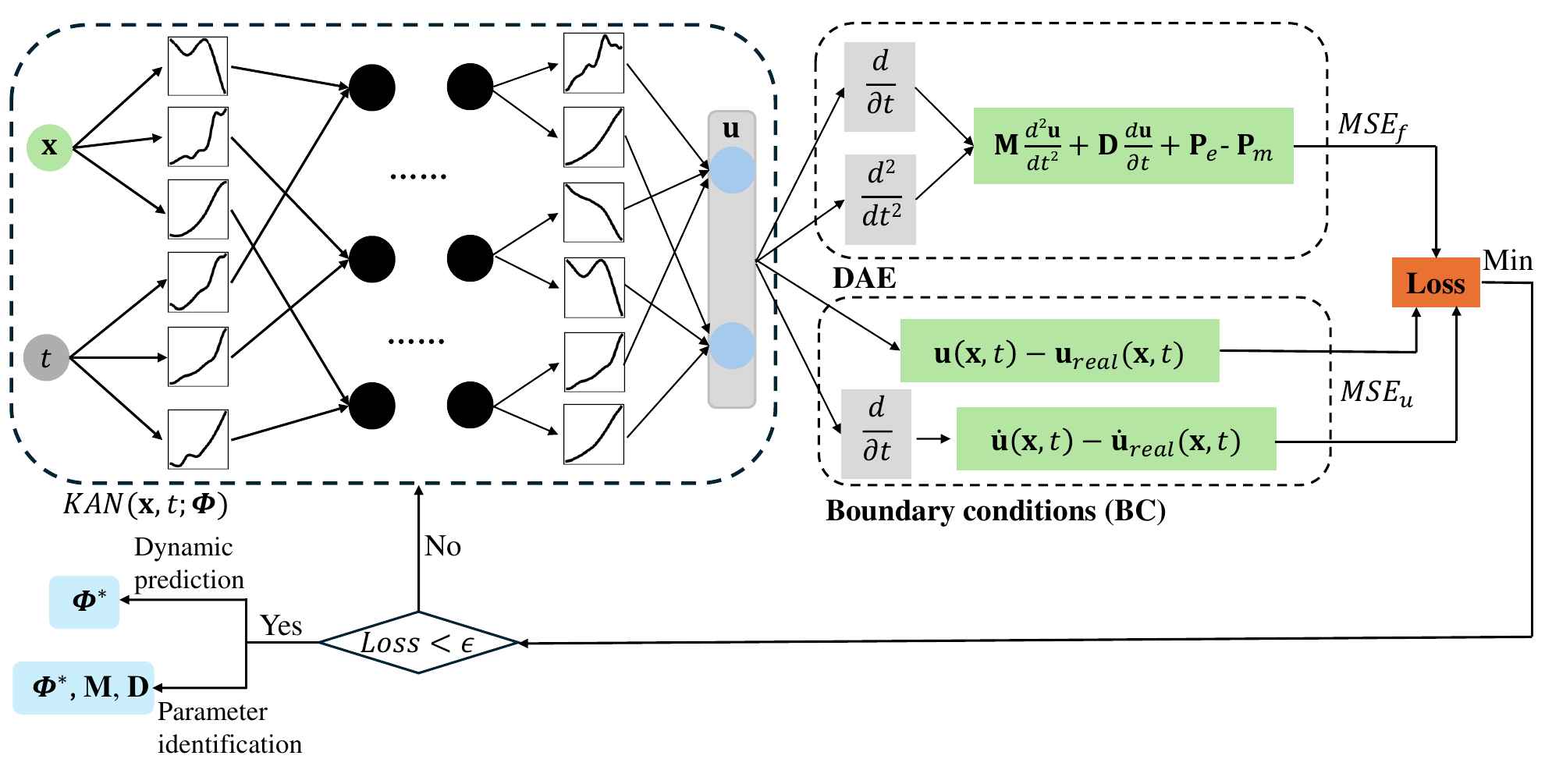}
    \caption{Physics-Informed Kolmogorov-Arnold Network (PIKAN) for power system dynamics.}
    \label{fig:PINN_KAN}
\vspace{-1em}
\end{figure*}

To reduce the dependency on training data and enhance the accuracy of the learned model in the PINNs-based power system dynamic model, we designed the PIKAN, as shown in Fig. \ref{fig:PINN_KAN}.
The primary difference from the traditional PINN is that we utilize KAN to predict $\textbf{u}(t, \textbf{x})$ based on the input state $\textbf{x}$ and time $t$.
This PIKAN offers two advantages: 1) increased model learning accuracy, and 2) reduced network size without sacrificing accuracy, which will be demonstrated in Section IV.

\textit{1) PIKAN for capturing power system
dynamics}: When the PIKAN is used for capturing power system dynamics, we assume system parameters \boldsymbol{$\lambda = [\textbf{M}; \textbf{D}; \textbf{B}]$} in equation (\ref{eq5}) are known.
Therefore, the input of KAN is defined as $\textbf{x}: = \textbf{P}_\textbf{m}$.
By inputting $\textbf{P}_\textbf{m}$ and time period of interest to the PIKAN in Fig. \ref{fig:PINN_KAN}, it can predict the voltage angle of each bus, i.e., $\textbf{u} = \boldsymbol{\theta} (t, \textbf{P}_\textbf{m})$.
The output of KAN is fed into the DAE module of the PIKAN to incorporate the power system dynamics model, as described by equation (\ref{eq5}), into the neural network architecture.
The training objective is to optimize the activation function $\boldsymbol{\Phi}$ to minimize loss function in equation (\ref{eq:16a}) or (\ref{eq:16b}).
Thus, by minimising the total loss function over the KAN parameters, we can obtain the optimal KAN:
\begin{equation}\label{eq:16c}
\begin{aligned}
\boldsymbol{\Phi}^* = \arg\min_{\boldsymbol{\Phi}} (MSE_u + MSE_f)
\end{aligned}
\end{equation}
Solving the above highly non-convex and  multi-parameter optimization problem is challenge. We can use the LBFGS or Adam optimiser to get a solution.
We refer to the PIKAN using the loss function in equation (\ref{eq:16a}) as PIKAN-I, and the PIKAN using the loss function in equation (\ref{eq:16b}) as PIKAN-II.
In other words, PIKAN-I uses only the measurements of the voltage angle $\boldsymbol{\theta}$ to train the KAN, while PIKAN-II uses both the voltage angle $\boldsymbol{\theta}$ and the angular frequency $\omega$ measurements to train the KAN.
The proposed PIKAN for power system dynamics can be summarized in \textbf{Algorithm 1}.

\textit{2) PIKAN for power system
parameter identification}: Estimating power system inertia and damping coefficients is crucial for maintaining frequency stability. With the increased installation of inverter-based resources (IBRs) in modern power systems, the inertia and damping constants can vary with the control strategies employed, potentially affecting system stability and dynamic performance. Therefore, it is necessary to frequently estimate these parameters.
When the PIKAN is used for parameter identification, $\textbf{M}$ and $\textbf{D}$ in \boldsymbol{$\lambda$} will be unknown in equation (\ref{eq5}).
The structure of the KAN remains unchanged, except that the $\textbf{M}$ and $\textbf{D}$ parameters are now considered as additional variables during the minimization of the loss function in the network training process.
So, by minimising the total loss function over the KAN parameters and power system uncertain parameters, we can obtain the optimal KAN:
\begin{equation}\label{eq:16d}
\begin{aligned}
\boldsymbol{\Phi}^*, \textbf{M}^*, \textbf{D}^* = \arg\min_{\boldsymbol{\Phi}, \textbf{M}, \textbf{D}} (MSE_u + MSE_f)
\end{aligned}
\end{equation}
The proposed PIKAN for power system parameter identification can be summarized in \textbf{Algorithm 2} (see Appendix).

\begin{algorithm}
\caption{PIKAN for capturing power system dynamics}\label{alg:three}
\KwData{Power system training and test dataset generated by time domain simulation; Power system parameters (e.g., $\textbf{M}$, $\textbf{D}$, and $\textbf{B}$)}
\KwResult{KAN parameters}
 Initialize KAN parameters: $\{\boldsymbol{\Phi}_{l}\}_{l=1}^L$, $G$, and $k_b$\;
 Specify the loss function as equation (\ref{eq:16a}) or (\ref{eq:16b})\;
 Specify the initial \& boundary training data points: $\{(t_u^n, \textbf{x}_u^n), \textbf{u}^n\}_{n=1}^{N_u}$, and specify collocation training points: $\{(t_f^n, \textbf{x}_f^n)\}_{n=1}^{N_f}$\;
 Specify the test points: $\{(t_{test}^n, \textbf{x}_{test}^n), \textbf{u}_{test}^n\}_{n=1}^{N_{test}}$\;
 Set the maximum number of training steps $N$, and learning rate\;
 \While{$n_{iter} < N$}{
  Forward pass of KAN to calculate all $\textbf{u}(t_u^n, \textbf{x}_u^n)$. If loss function (\ref{eq:16b}) is adopted, further calculate $\dot{\textbf{u}}(t_u^n, \textbf{x}_u^n)$ using automatic differentiation\;
  Calculate $MSE_u$ based on the output of KAN and the measurements\;
  Calculate $MSE_f$ based on the output of KAN and the power system dynamics given in equation (\ref{eq5})\;
  Find the best KAN parameters to minimize the loss function using the LBFGS optimizer\;
  \If{$n_{iter}$ \% 10 == 0}
{Evaluate the performance of the PIKAN agent over the test points based on equation (\ref{eq:17})\;}
 }
\end{algorithm}

To measure the performance during the training, we defined the mean squared error (MSE) of the predictions on the test dataset as
\begin{equation}\label{eq:17}
MSE_t = \frac{1}{N_{test}} \sum_{n=1}^{N_{test}} |\boldsymbol{\theta}_{\text{pred},n} - \boldsymbol{\theta}_n|^2 
\end{equation}
where $n$ is the index of the sampled test data point. 
$\boldsymbol{\theta}_{pred, n}$ and $\boldsymbol{\theta_n}$ are the predicted and real voltage angle vector of all the buses in the system, respectively.
$N_{test}$ is the total points of the test dataset.

To evaluate the predictive performance of the well-trained PIKANs, we defined the relative prediction error of the voltage angle as:
\begin{equation}\label{eq:18}
\text{e}_{\theta} = \frac{\| \boldsymbol{\theta}^{0:T} - \boldsymbol{\theta}_{pred}^{0:T} \|_2}{\| \boldsymbol{\theta}^{0:T} \|_2} \\
= \frac{\sqrt{\sum_{i=1}^{n_b} \sum_{t=0}^{T} (\theta_{i}^t - {\theta}_{pred, i}^t)^2}}{\sqrt{\sum_{i=1}^{n_b} \sum_{t=0}^{T} (\theta_{i}^t)^2}}
\end{equation}
where $\boldsymbol{\theta}^{0:T
}$ and $\boldsymbol{\theta}_{pred}^{0:T
}$ represent the actual and predicted voltage angles of all buses from time 0 to $T$, respectively.
$\theta_{i}^t$ and $\theta_{pred, i}^t$ are the actual and predicted voltage angle of bus $i$ at time $t$, respectively.
$\|\cdot\|$ is the $l^2$ norm for finite-dimensional vectors.
For the inertia and damping coefficients identification performance, we defined the relative estimation error as:
\begin{equation}\label{eq:19}
\begin{aligned}
\text{e}_{M_i} = \frac{|M_i - M_{pred, i}|}{M_i} , \
\text{e}_{D_i} = \frac{|D_i - D_{pred, i}|}{D_i}
\end{aligned}
\end{equation}
where $M_i$ and $M_{pred, i}$ represent the actual and predicted inertia coefficients of the generator connected to bus $i$, respectively.
$D_i$ and $D_{pred, i}$ represent the actual and predicted damping coefficients of bus $i$, respectively.

\section{Simulation and Results}
\begin{figure}
    \centering
    \includegraphics[width=80mm]{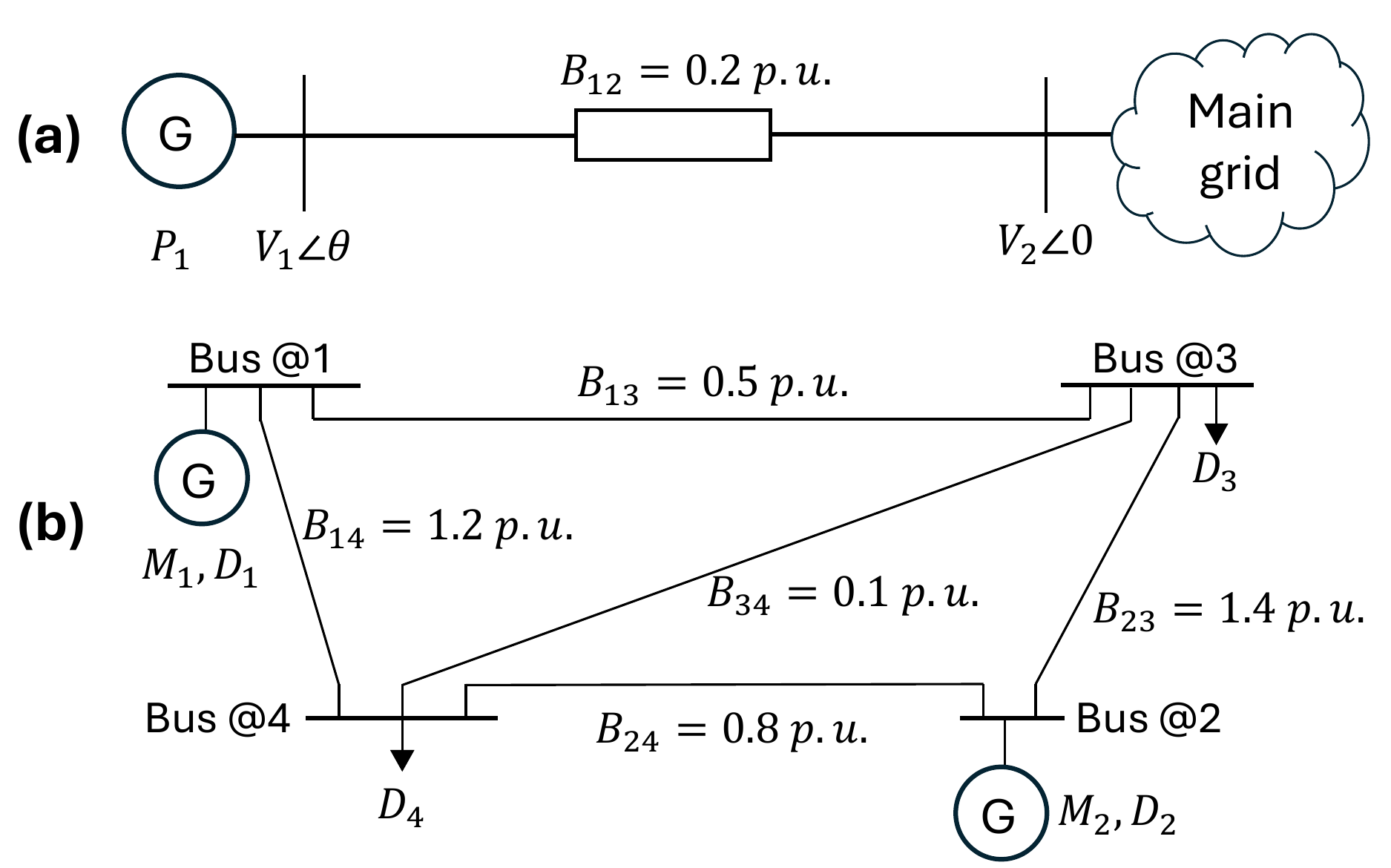}
    \caption{Testing systems: (a) SMIB power system, (b)  4-bus system with two generator.}
    \label{fig:Testing_system}
\vspace{-1em}
\end{figure}

The performance of the proposed PIKANs for frequency dynamics was demonstrated on a SMIB power system and a 4-bus 2-generator system, as shown in Fig. \ref{fig:Testing_system}. 
To generate the training and test datasets, we utilized time domain simulations implemented with SciPy in Python.
The generated frequency dynamic data is with a time step of 0.1$s$ over time window [0, $T$] for each trajectory.
The testing power system parameters are presented in Table 1 and Fig. \ref{fig:Testing_system}.
In the SMIB system, we assume initial values for $\theta_1$ and $\omega_1$ to be 0.1 rad and 0.1 rad/s, respectively. The value of $P_{m_1}$ ranges between 0.08 p.u. and 0.18 p.u., within which the SMIB system remains stable. 
In this case setting, we generated 100 trajectories.
For each trajectory, the training and test datasets consist of time intervals from 0 to 20 seconds with a 0.1-second step, including the corresponding $\theta$ values at each time step and the corresponding power injection value $P_{m_1}$.
For the 4-bus 2-generator system, similar to the setup in reference \cite{stiasny2021physics}, we assume the system is in equilibrium at $t=0$. 
We then perturb the system with a constant input signal $\textbf{P}_\textbf{m} = a \times [0.1, 0.2, -0.1, -0.2]$ p.u. for $t > 0$ in each trajectory. 
We generated 19 trajectories, with $a$ ranging from 0.5 to 9.5 in increments of 0.5.
For each trajectory, the training and test datasets consist of time intervals from 0 to 5 seconds with a 0.1-second step, including the corresponding [$\theta_1$, $\theta_2$, $\theta_3$, $\theta_4$] values at each time step and the corresponding input signal $\textbf{P}_\textbf{m}$.
We conducted PIKANs training and performance testing in PyTorch on an Intel Xeon(R) Gold 6248R CPU @3.00 GHz × 48 Windows based server with 64 GB RAM.

\begin{table}[h!]
\caption{Parameters of the case studies}
\label{table:1}
    \centering
\begin{tabular}{c *{6}{S[table-format=2.3]} }
    \toprule
\multirow{2.4}{*}{Parameters}
    &   \multicolumn{2}{c}{SMIB system}
        &   \multicolumn{2}{c}{4-bus 2-generator system}    \\
    \cmidrule(lr){2-3}
    \cmidrule(l){4-5}
    & {$M$ (p.u.)} & {$D$ (p.u.)}     
    & {$M$ (p.u.)}  & {$D$ (p.u.)}    \\
    \midrule
 Bus @1  & 0.4  & 0.15 & 0.3   & 0.15   \\
 Bus @2  & \text{---} & \text{---} & 0.2  &   0.3      \\
 Bus @3  & \text{---}  & \text{---} & 0   & 0.25  \\
 Bus @4  & \text{---}  & \text{---} & 0   & 0.2  \\
    \bottomrule
\end{tabular}
\begin{tablenotes}
      \small
      \item Note: The line parameters of the testing systems can be found in Fig. \ref{fig:Testing_system}.
    \end{tablenotes}
\vspace{-1em}
\end{table}

\subsection{Data-driven solution of frequency dynamics}
In the study of capturing frequency dynamics, the inertia and damping coefficients of the testing systems are known parameters.
We evaulated the capability of the PIKANs to accurately predict trajectories of $\boldsymbol{\theta}$ and $\boldsymbol{\omega}$ for uncertain power injections.

\begin{figure}
    \centering
    \includegraphics[width=90mm]{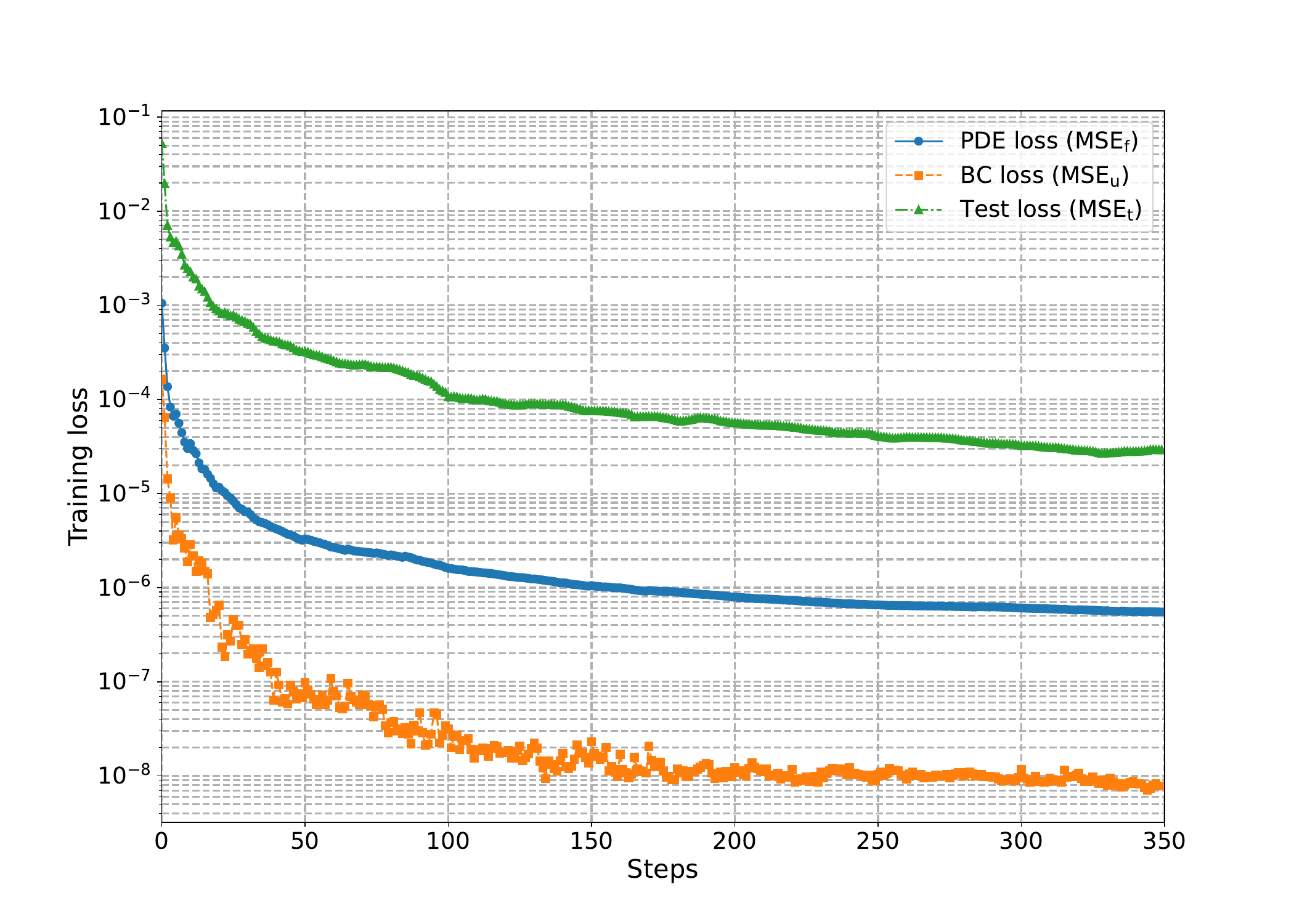}
    \caption{Training convergence process of the PIKAN-I algorithm for capturing SMIB system frequency dynamics. The LBFGS optimizer was employed, with parameter maximum iteration set to 20.
    Thus, each optimization step in the figure contains 20 iterations.}
    \label{fig:loss_case_I}
\vspace{-1em}
\end{figure}

\begin{figure}[t]
    \centering
    \includegraphics[width=90mm]{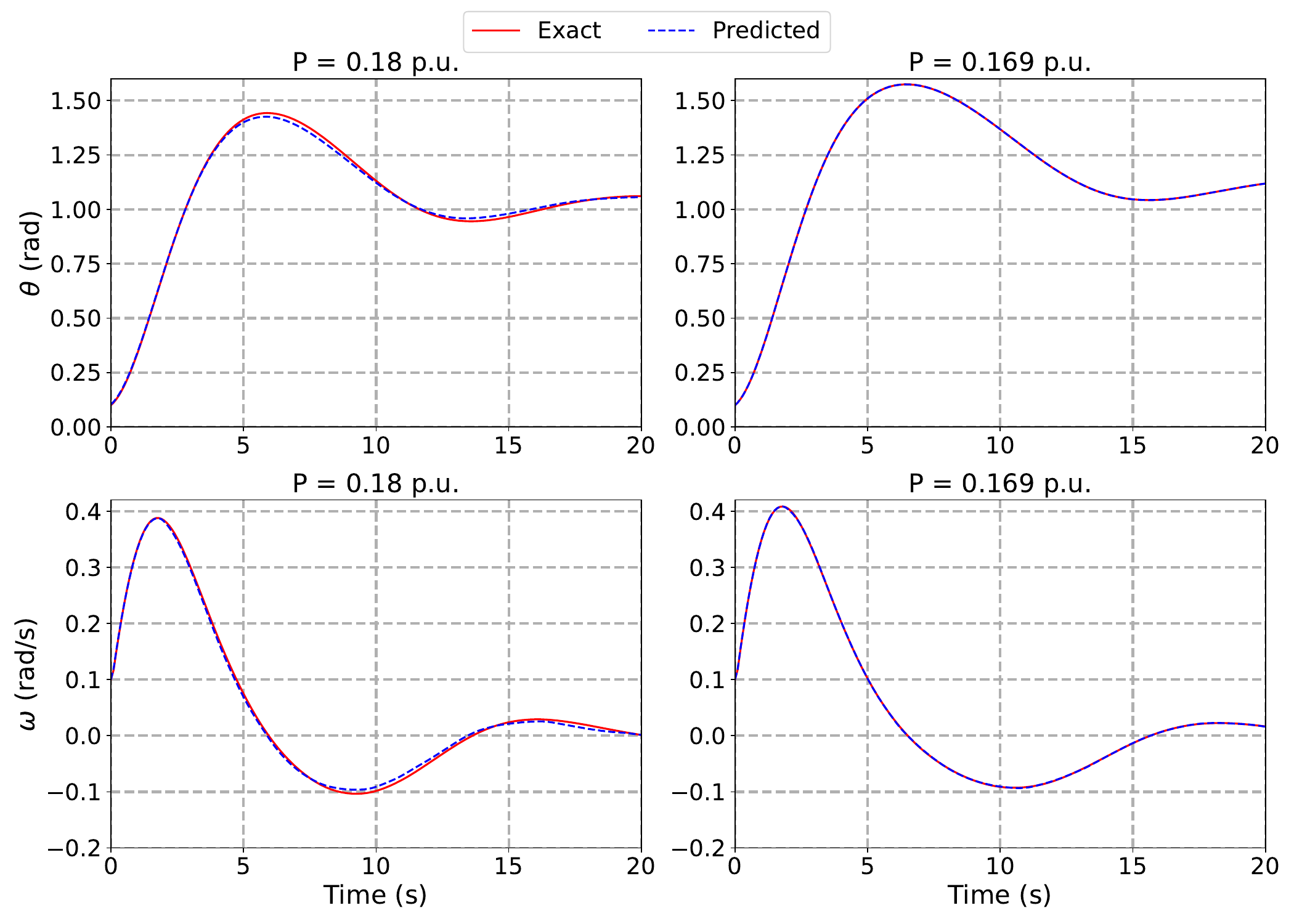}
    \caption{Comparison of the predicted and exact solution for the voltage angle and frequency with the PIKAN-I for SMIB power system dynamics.}
    \label{fig:fit_Case_I}
\vspace{-1em}
\end{figure}

1) \textit{SMIB system:} For the SMIB system, we used a 2-layer KAN with a shape of [2, 5, 1].
In each training step, the randomly sampled time $t$ and power injection $P_{m_1}$ were fed into the KAN, and trained to minimize the loss function in equation (\ref{eq:16a}) for the PIKAN-I algorithm (or equation (\ref{eq:16b}) for the PIKAN-II algorithm).
For both PIKAN-I and PIKAN-II algorithms, the intervals of the B-spline were set to $G=10$, and the order of the B-spline was set to $k_b=3$.
We set $N_u$ = 40, $N_f$ = 800, and $N_{test} = 20,100$.
The training convergence process of the PIKAN-I algorithm is depicted in Fig. \ref{fig:loss_case_I}.
It shows that the PIKAN-I converges quickly and achieves lower losses within hundreds of training steps.
Fig. \ref{fig:fit_Case_I} depicts the comparison between the PIKAN-I predicted and the actual trajectory of the angle $\theta$ and the angular frequency $\omega$ of bus 1 in the SMIB system.
The angular frequency $\omega$ in the figure was calculated by differentiating the signal associated with the voltage angle $\theta$.
In the left figures of Fig. \ref{fig:fit_Case_I}, we present the  least accurate estimation of the voltage angle and frequency trajectory, yielding a relative prediction error ($\text{e}_{\theta}$) of 1.06\%.
Conversely, in the right figures, we demonstrate the most accurate estimation of the voltage angle and frequency trajectory, achieving a relative prediction error ($\text{e}_{\theta}$) of 0.014\%.
The median value of the prediction error on voltage angle over the 100 trajectories is 0.688\%, which indicate that the PIKAN-I is able to predict the trajectory of the angle with high accuracy.

\begin{figure}
    \centering
    \includegraphics[width=90mm]{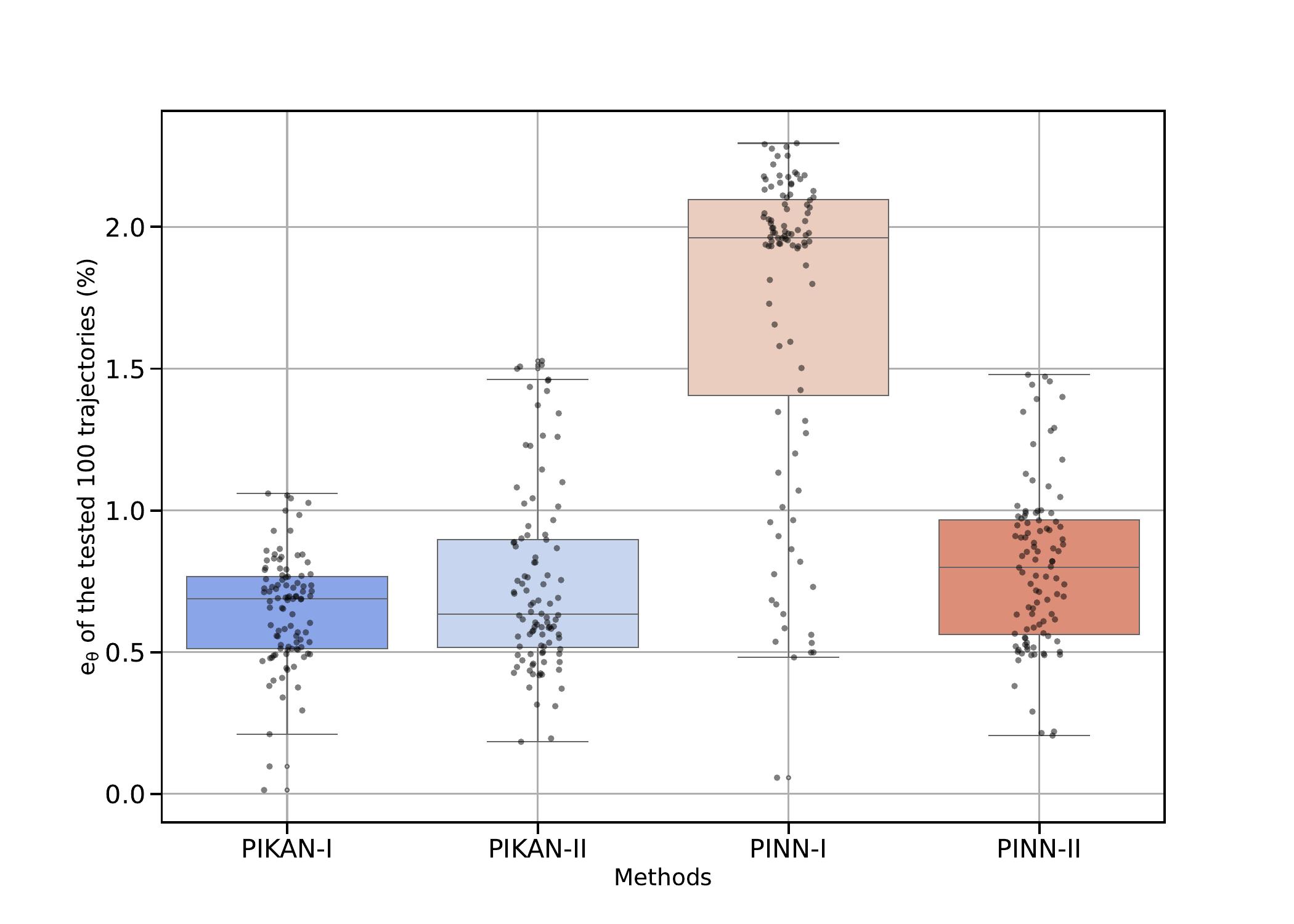}
    \caption{Performance of the proposed method and the MLPs based PINNs for the SMIB system. The parameters and hyperparameters setting is same with reference \cite{misyris2020physics}. For the PINN-I and PINN-II, we set $N_u = 40$ and $N_f = 8,000$.}
    \label{fig:Figure_boxplot}
\vspace{-1em}
\end{figure}

If we use measurements of both $\theta$ and $\omega$ to train the KAN, denoted as the PIKAN-II algorithm, the accuracy of the agent can be further improved, with the median value of the prediction error on the voltage angle decreasing to 0.633\% (see Table II).
We also compared the performance of the proposed method with the MLP-based PINNs for power systems proposed in \cite{misyris2020physics} and \cite{stiasny2021physics}. The prediction errors for the 100 tested trajectories are presented in Fig. \ref{fig:Figure_boxplot} and Table II.
The proposed method outperforms the traditional PINNs, demonstrating the effectiveness of the PIKANs in learning the dynamics of SMIB systems.
From the results in Fig. \ref{fig:Figure_boxplot}, we can observe that incorporating measurements of $\omega$ (i.e., using the loss function defined in equation (\ref{eq:16b})) during training improves the performance of the agent for both the PIKAN and traditional PINN methods.

\begin{table*}[h!]
\caption{Dynamic capturing study results: Estimation error of
 the trajectory of $\theta$($t$)}
\label{table:2}
    \centering
\begin{tabular}{c *{6}{S[table-format=2.3]} }
    \toprule
\multirow{2.4}{*}{Estimation error}
    &   \multicolumn{3}{c}{SMIB system}
        &   \multicolumn{3}{c}{4-bus 2-generator system}    \\
    \cmidrule(lr){2-4}
    \cmidrule(l){5-7}
    & {$Max$ (\%)} & {$Min$ (\%)} & {$Median$ (\%)}     
    & {$Max$ (\%)} & {$Min$ (\%)} & {$Median$ (\%)}    \\
    \midrule
 PIKAN-I  & 1.06  & 0.014 & 0.688   & 4.85 & 0.043 & 4.64    \\
 PIKAN-II  & 1.53  & 0.184 & 0.633   & 1.94 & 0.040 & 0.538    \\
PINN-I (\cite{misyris2020physics})  & 2.30  & 0.057 & 1.96   & 6.35 & 0.151 & 5.03   \\
PINN-II (\cite{stiasny2021physics})  & 1.48  & 0.206 & 0.800   &5.98  &0.076  &2.59    \\
 
    \bottomrule
\end{tabular}
\vspace{-1em}
\end{table*}

2) \textit{4-bus 2-generator system:} 
To further test the performance of the proposed method in capturing the dynamics of multi-machine power systems, we evaluated it on a 4-bus 2-generator system as shown in Fig. \ref{fig:Testing_system} (b).
For this case study, we employed a 2-layer KAN with a structure of [5, 10, 4].
In each training step, the randomly sampled time $t$ and power injection [$P_{m_1}$, $P_{m_2}$, $P_{m_3}$, $P_{m_4}$] are fed into the KAN, which is then trained to minimize the loss function in equation (\ref{eq:16a}) or (\ref{eq:16b}), ultimately outputting the voltage angles of the four buses at time $t$.
For both PIKAN-I and PIKAN-II, the intervals of the B-spline were set to $G=5$, and the order of the B-spline was set to $k_b=3$.
We set $N_u$ = 80, $N_f$ = 4000, and $N_{test} = 969$.

\begin{figure}
    \centering
    \includegraphics[width=92mm]{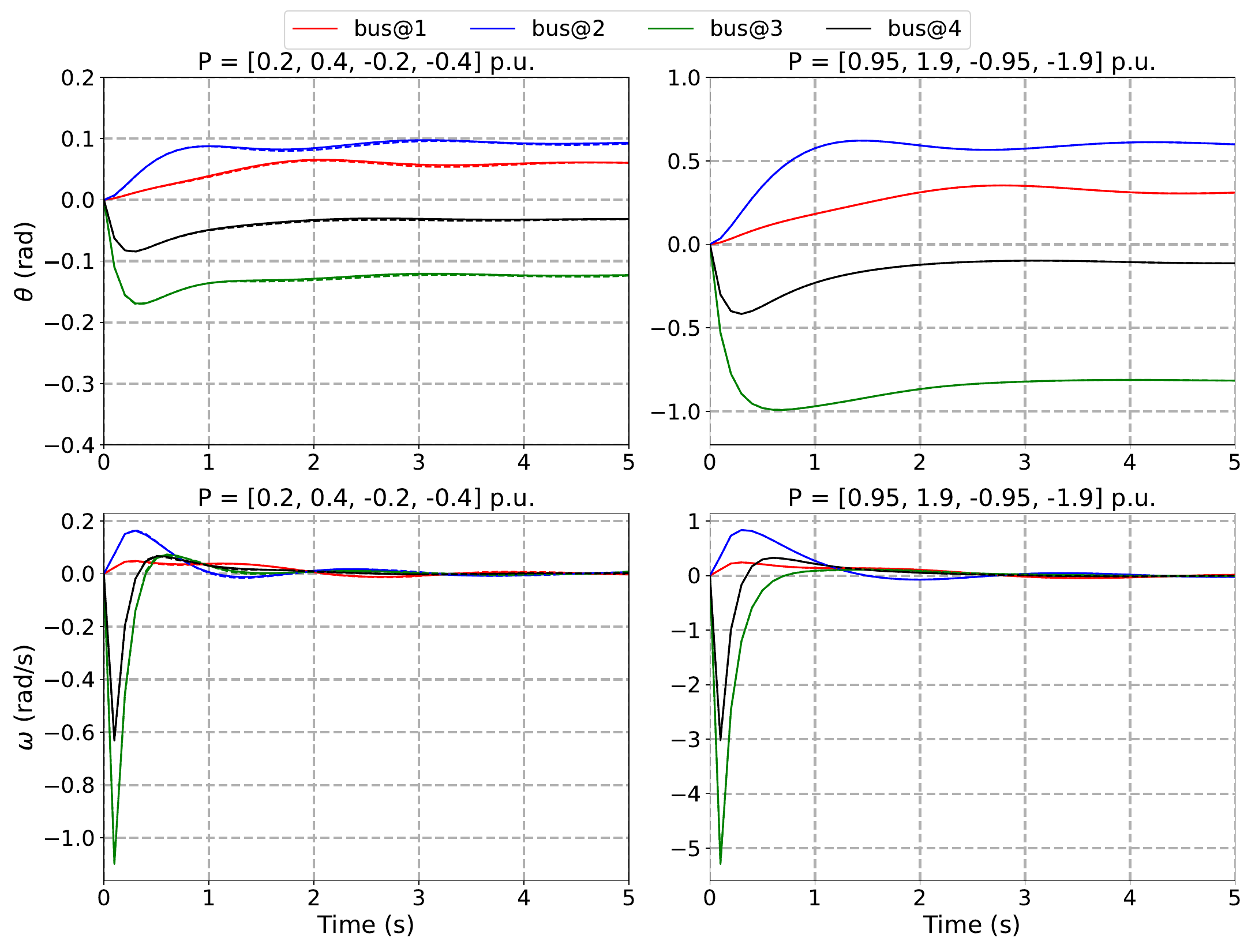}
    \caption{Comparison of the predicted and exact solutions for the voltage angle and frequency using PIKAN-II for the 4-bus 2-generator power system dynamics. Solid lines represent the exact trajectory, while dashed lines represent the predicted trajectory by PIKAN-II.}
    \label{fig:PINN_KAN_4_bus_curve}
\vspace{-1em}
\end{figure}

Fig. \ref{fig:PINN_KAN_4_bus_curve} depicts the comparison between the predicted and the actual trajectory of the angle [$\theta_1$, $\theta_2$, $\theta_3$, $\theta_4$] and the frequency [$\omega_1$, $\omega_2$, $\omega_3$, $\omega_4$] of 4 buses in the system.
In the left figures of Fig. \ref{fig:PINN_KAN_4_bus_curve}, we present the least accurate estimation of the voltage angle and frequency trajectory, yielding a relative prediction error ($\text{e}_{\theta}$) of 1.94\%.
Conversely, in the right figures, we demonstrate the most accurate estimation of the voltage angle and frequency trajectory, achieving a relative prediction error ($\text{e}_{\theta}$) of 0.04\%.
The median value of the estimation error on voltage angle over the 19 trajectories is 0.538\%, indicating that PIKAN-II can predict the trajectory of the angle with high accuracy.
In contrast, the traditional PINN-I and PINN-II algorithms performed much worse, with median estimation errors on the voltage angle of 5.03\% and 2.59\%, respectively.
The performance comparisons between PIKANs and PINNs are summarized in Table II. The results on the 4-bus 2-generator system also demonstrate that the proposed method outperforms traditional PINN-based approaches.

\subsection{Data-driven inertia and damping coefficients identification}
In the parameter identification study, the inertia and damping coefficients of the testing systems are unknown parameters. 
We assessed the capability of PIKANs to accurately estimate these unknown parameters from observed trajectories.

\begin{figure}
    \centering
    \includegraphics[width=83mm]{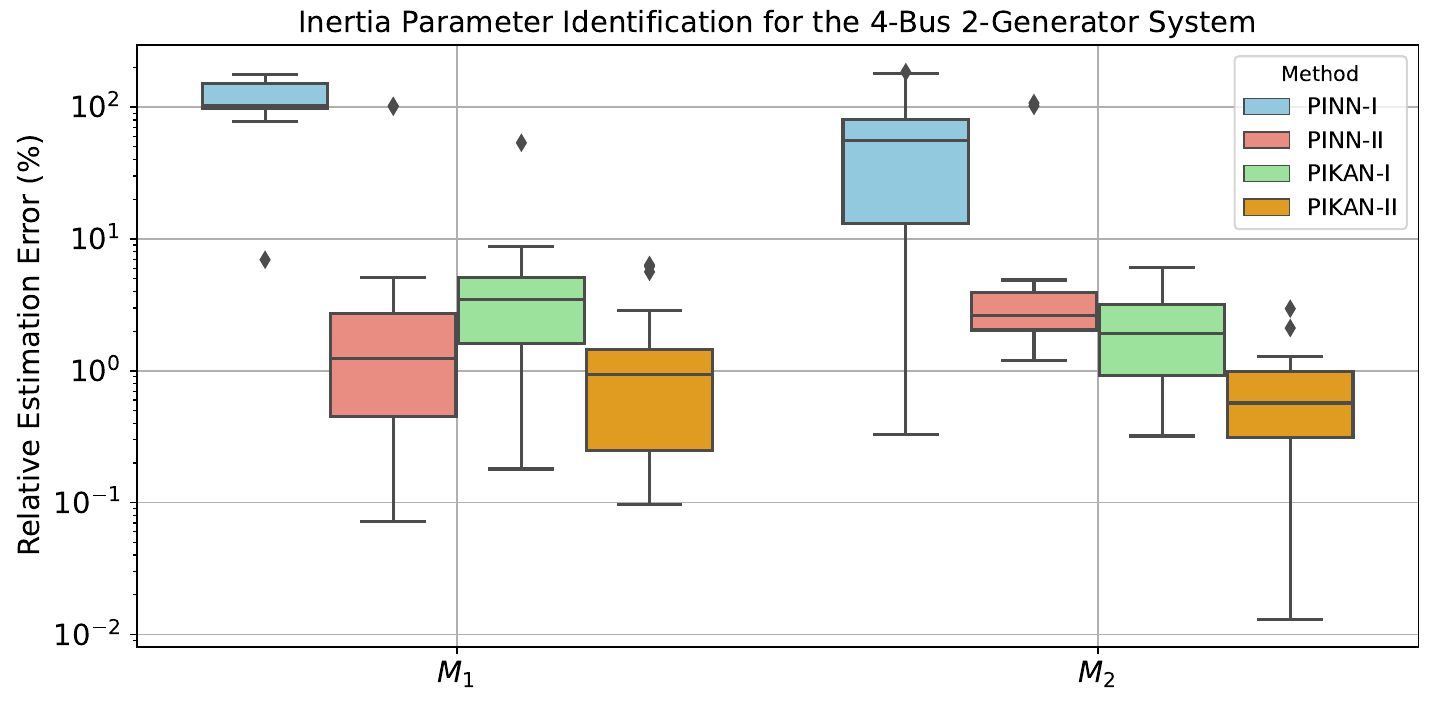}
    \caption{Inertia coefficients estimation errors of PIKANs and PINNs.}
    \label{fig:4_bus_identification_m}
\end{figure}

\begin{figure}
    \centering
    \includegraphics[width=83mm]{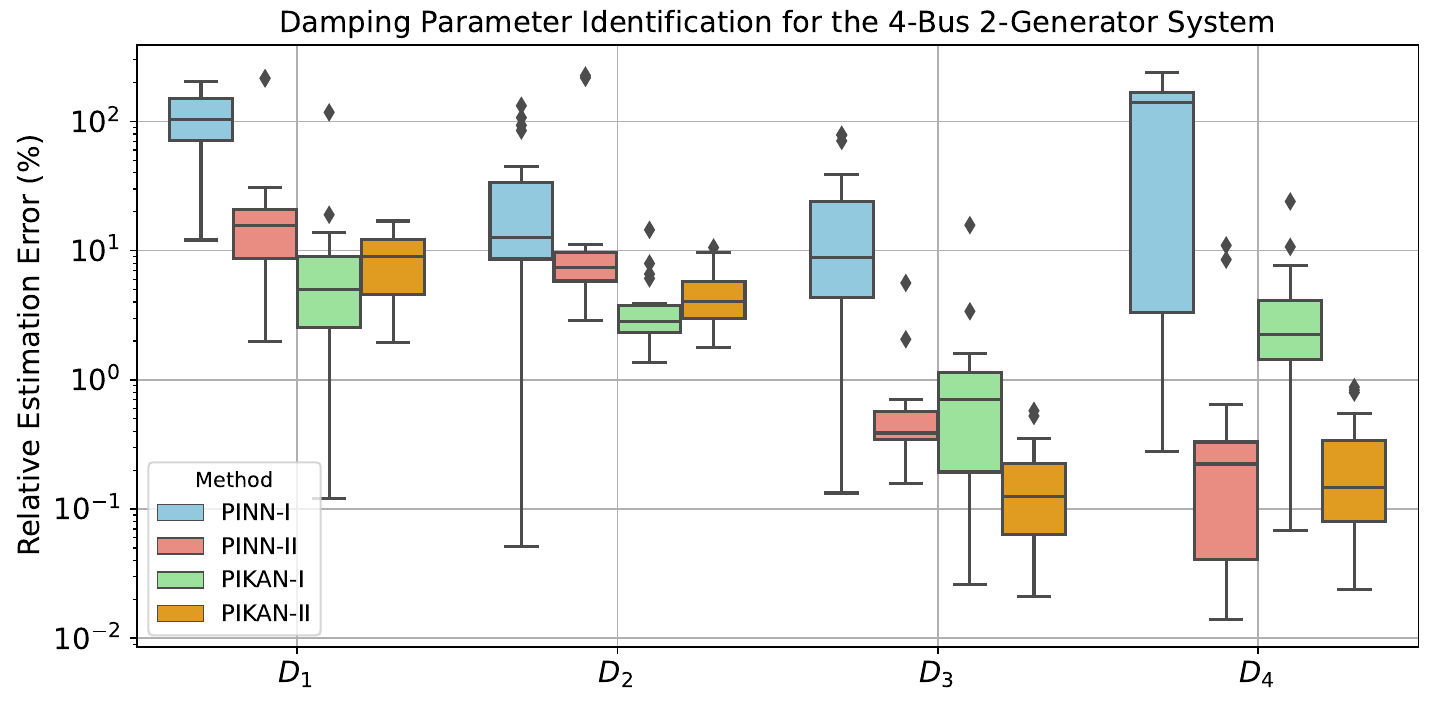}
    \caption{Damping coefficients estimation errors of PIKANs and PINNs.}
    \label{fig:4_bus_identification_d}
\vspace{-1em}
\end{figure}

The parameters and hyperparameters of the PIKANs for assessing inertia and damping coefficients are the same as those of the PIKANs agents in Section IV-A.
Since the neural network's weights are initialized randomly, we run each estimation of the four algorithms 20 times.
Figs. \ref{fig:4_bus_identification_m} and \ref{fig:4_bus_identification_d} show the distribution of parameter estimation errors on the 4-bus 2-generator system for the proposed method and the comparison methods.
PIKAN-I achieves a median relative error below 10\% for evaluating the inertia and damping coefficients of the system. In contrast, PIKAN-II demonstrates significantly better performance, achieving a median relative error of around 1\% for inertia coefficients and 0.1\% for several damping coefficients.
The traditional PINNs, however, perform much worse than the proposed methods. Additionally, we observed that incorporating measurements of $\omega$ during training can significantly improve the parameter estimation accuracy of the agent for both the PIKAN and traditional PINN methods.

\subsection{Number of network parameters vs. PIKAN performance}
Results in Tables \ref{table:2} and \ref{table:3} indicate that, for the SMIB case, PIKANs achieved greater accuracy in grid dynamic learning while using only 41\% of the network size of the PINNs. Similarly, for the 4-bus 2-generator case, PIKANs achieved higher accuracy while utilizing only 58\% of the network size of the PINNs.

\begin{figure}
    \centering
    \includegraphics[width=83mm]{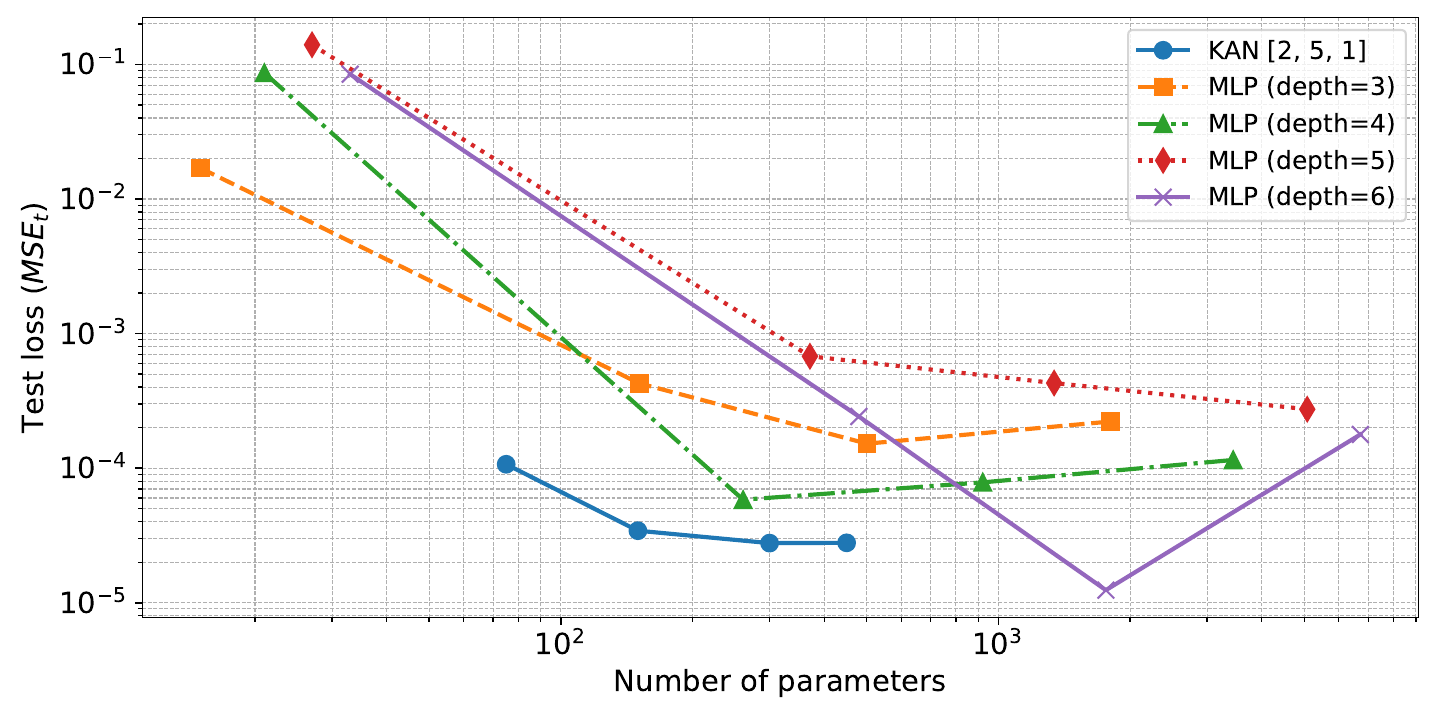}
    \caption{Scaling laws of losses against the number of parameters for different physics-informed neural networks applied to the SMIB system.}
    \label{fig:scaling_law}
\end{figure}

For a $L$-layer KAN with layers of equal width $N$, there are in total $O(N^2LG)$ parameters, and an MLP only needs $O(N^2L)$ parameters for the same number of layers and width.
However, KANs typically achieve similar performance with a much smaller $N$ compared to MLPs.
Fig. \ref{fig:scaling_law} illustrates the scaling laws of losses as a function of the number of parameters in both PIKANs and PINNs. 
The results demonstrate that KANs exhibit steeper scaling laws than MLPs. 
This implies that PIKANs can achieve comparable or even superior accuracy in power system dynamic learning with fewer parameters than PINNs.
The implications of these findings are significant. They suggest that while KANs may initially seem to require more parameters due to the $O(N^2LG)$ scaling, their ability to use a smaller $N$ effectively reduces the overall parameter count needed for high performance. Consequently, PIKANs offer a more efficient and scalable solution for complex learning tasks in power systems, surpassing traditional MLP-based PINNs in terms of both accuracy and parameter efficiency.

\subsection{Reduced data dependency}
\begin{figure}
    \centering
\includegraphics[width=82mm]{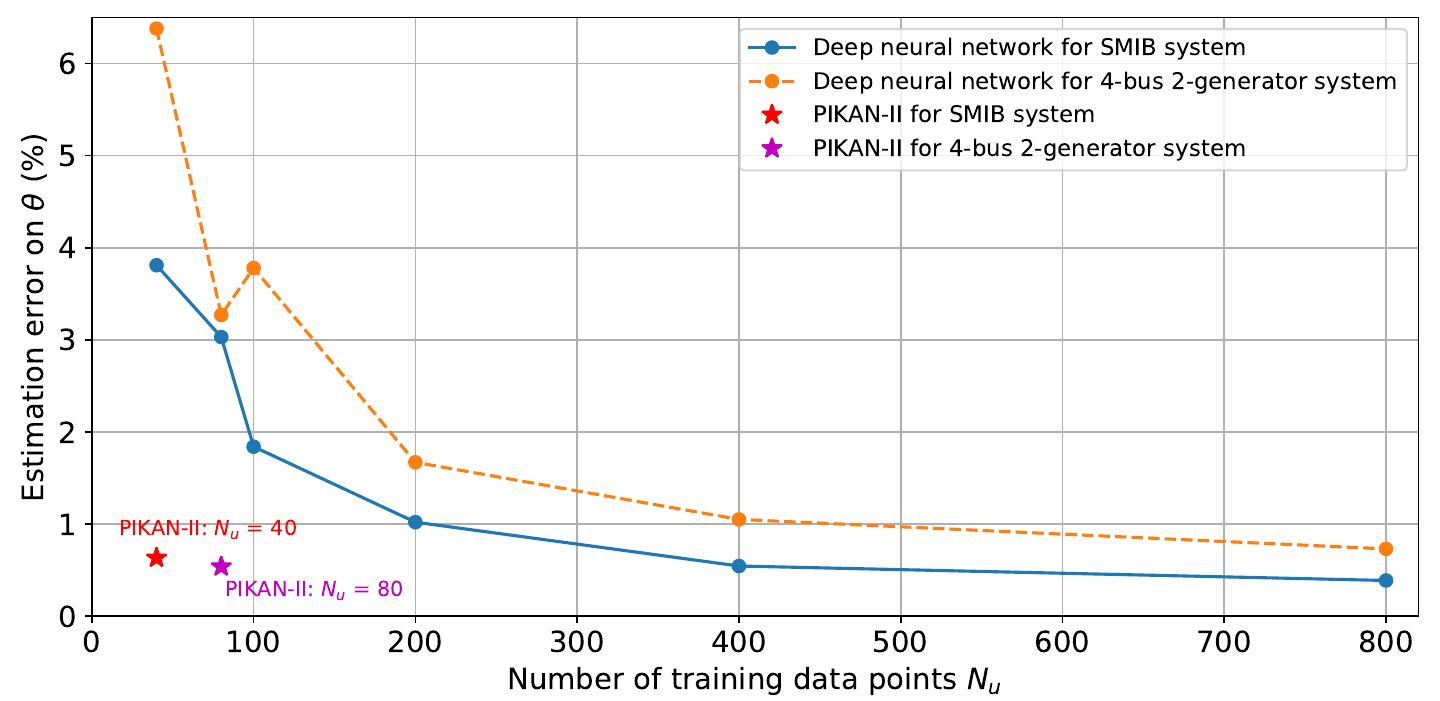}
    \caption{Performance comparison of traditional DNNs and PIKANs with varying numbers of training data points. For the SMIB case, a [2, 10, 10, 10, 10, 10, 1] DNN architecture is employed. For the 4-bus system, a [5, 30, 30, 4] DNN architecture is utilized, and the Adam optimizer was employed.}
    \label{fig:NN_training_points}
\vspace{-1em}
\end{figure}

PIKANs introduce a KAN training framework designed to leverage the inherent dynamics of power systems. Consequently, compared to traditional DNNs without a physics-informed architecture, PIKANs can significantly reduce the required size of the training dataset.
For the two testing systems in Table II, the performance of traditional DNNs, employing identical architecture and parameters as the PINNs, varies with the number of training data points, as illustrated in Fig. \ref{fig:NN_training_points}.
From the results, it is observed that PIKANs can achieve similar or even better performance while requiring only 10\% of the training data points compared to traditional DNNs.

\section{Discussion}
This paper introduces, for the first time in power systems, a KAN-based PINN (i.e., PIKAN) approach that explicitly considers the swing equations describing the frequency behavior of grids. 
As a promising alternative to traditional MLPs, the proposed PIKANs for power system dynamics can achieve comparable or even higher accuracy with fewer neural network parameters compared to MLP-based PINNs.
The advantage of PIKANs is particularly significant given the challenges of training large neural networks, such as large language models (LLMs), which are resource-intensive and consume substantial amounts of energy.
This opens up numerous opportunities in power systems, as PIKANs can potentially be used to accurately and efficiently solve DAEs in power grids.
In addition, PIKANs require only a very limited amount of training data. 
For instance, for the SMIB system, PIKANs need only $N_u=40$ points $\{(t_u^n, \textbf{x}_u^n), \textbf{u}^n\}_{n=1}^{N_u}\}$ to train the agent.
Even for a larger power system, such as the 4-bus 2-generator system, PIKANs still need only $N_u=80$ training data points.
Although we require significantly more collocation points (for example, $N_f = 800$ for the SMIB case) to evaluate the $MSE_f$ term in the loss function given in equation (\ref{eq:16a}), it is important to note that this evaluation is not dependent on measured voltage angle and angular frequency data. 
This means we can generate any number of collocation points without needing to produce labels for those data points.

Similar with traditional PINNs, PIKANs have the capability to directly compute the voltage angle at any given time step $t_1$.
In contrast, numerical methods must integrate starting from the initial conditions at $t = t_0$ and proceed sequentially to reach $t = t_1$. 
This provides significant advantages over traditional numerical integration methods.

In this study, our primary focus was on exploring how PIKANs could achieve higher accuracy in learning power system dynamics while maintaining a smaller network size.
Theoretically, KANs outperform MLPs in terms of accuracy, interpretability, and reduction of catastrophic forgetting.
Nevertheless, to fully harness the potential of PIKANs, several challenges must be addressed.
\begin{enumerate}
    \item \textit{Training and computing time}: From the results presented in Table \ref{table:3}, it is evident that the training of PIKANs requires considerably more time compared to PINNs. 
    Liu et al. \cite{liu2024kan} attribute this slower training to the inefficiency of current activation functions in batch computations. 
Despite the extended training duration, the superior performance and accuracy of PIKANs may justify the additional time investment, especially in scenarios requiring high precision.
After offline training, we evaluate the PIKAN's performance based on its online computational speed required to solve the DAE defined by equation (\ref{eq5}).
For 19 different initial conditions of the 4-bus 2-generator system, the ode45 solver averages 0.017 seconds to solve the swing equations across the time interval from 0 seconds to 5 seconds, whereas PIKAN averages 0.024 seconds.
In future research, we aim to explore techniques utilizing more efficient activation functions, such as Jacobi polynomials proposed by \cite{SynodicMonth2024}, to substantially enhance training speeds.
And primary investigation in \cite{shukla2024comprehensive} demonstrates that Jacobi polynomials can reduce training times by two orders of magnitude compared to KANs using B-spline activation functions  in the context of solving specific PDEs.

\item \textit{Accuracy}:
Our simulation results demonstrated that KAN-based PINNs exhibit higher accuracy compared to MLP-based PINNs in modeling power system dynamics. Researchers have also found that KANs generally achieve greater accuracy than MLPs in most PDE problems \cite{wang2024kolmogorov}. However, whether KANs consistently outperform MLPs in various power dynamic problems requires further investigation. Additionally, understanding why PIKANs have higher accuracy than conventional PINNs warrants further exploration. One possible reason could be that KANs employ learnable activation functions, allowing for more complex learned activations compared to the fixed activation functions (such as ReLU) used in MLPs.

\item \textit{Interpretability}:  KANs have the potential to serve as foundational models for AI + Science due to their accuracy and interpretability \cite{liu2024kan}. With KANs, humans can interactively obtain the symbolic formula of the model's output, which significantly facilitates the analysis of complex physical systems, such as the dynamics of bulk power systems. 
However, in the case of the swing equations examined in this study, we observed that the symbolic formula provided by the well-trained PIKAN does not accurately capture the frequency dynamics of the two testing systems, despite the PIKAN model itself precisely predicting the grid dynamics.
This discrepancy may stem from the limited library of symbolic formulas available in the current version of KAN package in \cite{KANCode2024}, or perhaps the formula for the grid dynamics is not inherently symbolic.

\item \textit{Continual learning}: One drawback of MLPs is their tendency to forget previously learned tasks when transitioning from one task to another. Liu et al. \cite{liu2024kan} demonstrate that for a 1D regression task, KANs exhibit local plasticity and can prevent catastrophic forgetting by leveraging the locality of splines. However, the extent to which KANs can avoid catastrophic forgetting in more complex learning tasks, such as power system dynamics as explored in this study, remains unclear.
In our investigation, we observed that a well-trained PIKAN, initially trained on data from stable scenarios (e.g., $P_{m_1} \in [0.08, 0.18]$ p.u. for the SMIB case), tends to forget previously learned dynamics when further trained on dynamics from unstable scenarios (e.g., $P_{m_1} \in [0.20, 0.25]$ p.u. for the SMIB case). 
Therefore, further investigation into the continual learning capabilities of the proposed PIKANs is warranted in future research.
\end{enumerate}

\begin{table*}[h!]
\caption{Training time for results given in Table II}
\label{table:3}
    \centering
    \begin{tabular}{>{\centering\arraybackslash}m{2.0cm} >{\centering\arraybackslash}m{2.0cm} >{\centering\arraybackslash}m{2.0cm} >{\centering\arraybackslash}m{1.5cm} >{\centering\arraybackslash}m{1.5cm} >{\centering\arraybackslash}m{1.5cm} >{\centering\arraybackslash}m{1.5cm} >{\centering\arraybackslash}m{1.5cm}}
    \toprule
System & Methods & Network layers & \shortstack{Order of \\ B-spline ($k_b$)} & \shortstack{Intervals of \\ B-spline ($G$)} & \shortstack{No. of \\ parameters} & Training iterations & \shortstack{Training time \\ (ms/iter.)} \\
    \midrule
    \multirow{4}{*}{SMIB system} & PIKAN-I & {[2, 5, 1]} & 3 & 10 & 195 & 7000 & 87.5     \\
 & PIKAN-II  & {[2, 5, 1]}  & 3 & 10  & 195 & 7000 & 130     \\
& PINN-I (\cite{misyris2020physics})  & {[2, 10, 10, 10, 10, 10, 1]}  & \text{--} & \text{--}  & 481 & 50000 & 0.54  \\
& PINN-II (\cite{stiasny2021physics})  & {[2, 10, 10, 10, 10, 10, 1]}  & \text{--} & \text{--}  & 481 &10000  &3.41   \\
\midrule
    \multirow{4}{*}{\shortstack{4-Bus \\ 2-Generator System}} &  PIKAN-I  & {[5, 10, 4]}  & 3 & 5  & 720 & 3000 & 1225     \\
 & PIKAN-II  & {[5, 10, 4]}  & 3 & 5  & 720 & 3000 & 1390     \\
& PINN-I (\cite{misyris2020physics})  & {[5, 30, 30, 4]}  & \text{--} & \text{--}  & 1234 & 50000 & 2.89  \\
& PINN-II (\cite{stiasny2021physics})  & {[5, 30, 30, 4]}  & \text{--} & \text{--}  & 1234 &50000  &3.78   \\
    \bottomrule
\end{tabular}
\vspace{-1em}
\end{table*}

\vspace{-1em}
\section{Conclusions}
This is the first paper to propose physics-informed KANs for power system applications.
By integrating KAN with PINN, we achieve higher accuracy in solving the differential-algebraic equations of power systems with smaller neural network size compared to traditional MLP-based PINNs.
In our case studies, we showcased the effectiveness of the proposed PIKANs in accurately capturing the dynamics of power systems. 
Furthermore, we demonstrated their capability to identify uncertain system inertia and damping parameters, with high accuracy using a limited set of training data points. 
These results underscore the promising potential of the PIKANs for practical applications in power systems, opening up new avenues for their use.

\appendices
\section{}
See Algorithm 2.
\begin{algorithm}
\caption{PIKAN for grid parameter identification}\label{alg:three}
\KwData{Power system training and test dataset generated by time domain simulation}
\KwResult{KAN parameters and estimated inertia $\textbf{M}$ and damping $\textbf{D}$ parameters}
 Initialize KAN parameters: $\{\boldsymbol{\Phi}_{l}\}_{l=1}^L$, $G$, and $k_b$\;
 Initialize inertia $\textbf{M}$ and damping $\textbf{D}$ parameters\;
 Specify the loss function as equation (\ref{eq:16a}) or (\ref{eq:16b})\;
 Specify the initial \& boundary training data points: $\{(t_u^n, \textbf{x}_u^n), \textbf{u}^n\}_{n=1}^{N_u}$, and specify collocation training points: $\{(t_f^n, \textbf{x}_f^n)\}_{n=1}^{N_f}$\;
 Specify the test points: $\{(t_{test}^n, \textbf{x}_{test}^n), \textbf{u}_{test}^n\}_{n=1}^{N_{test}}$\;
 Set the maximum number of training steps $N$, and learning rate\;
 \While{$n_{iter} < N$}{
  Forward pass of KAN to calculate all $\textbf{u}(t_u^n, \textbf{x}_u^n)$. If loss function (\ref{eq:16b}) is adopted, further calculate $\dot{\textbf{u}}(t_u^n, \textbf{x}_u^n)$ using automatic differentiation\;
  Calculate $MSE_u$ based on the output of KAN and the measurements\;
  Calculate $MSE_f$ based on the output of KAN and the power system dynamics given in equation (\ref{eq5})\;
  Find the best KAN parameters and inertia $\textbf{M}$ and damping $\textbf{D}$ parameters to minimize the loss function using the LBFGS optimizer\;
  \If{$n_{iter}$ \% 10 == 0}
{Evaluate the performance of the PIKAN agent over the test points based on equation (\ref{eq:17})\;}
 }
\end{algorithm}
\vspace{-0em}

\ifCLASSOPTIONcaptionsoff
  \newpage
\fi

\bibliographystyle{IEEEtran}
\bibliography{IEEEabrv,KAN_Ref}

%
\end{document}